\documentclass[12pt]{iopart}
\usepackage{amsbsy}
\usepackage{graphicx, epsfig, amssymb} 
\usepackage{bm} 

\usepackage[linktocpage]{hyperref}
\usepackage[caption=false]{subfig}
\usepackage[usenames]{color}
\usepackage{url}
\usepackage{color}

\def\be{\begin{equation}}
\def\ee{\end{equation}}
\def\beq{\begin{eqnarray}}
\def\eeq{\end{eqnarray}}

\def\nn{\nonumber}

\begin{document}

\title{Black holes as particle detectors: evolution of superradiant instabilities}

\author{Richard Brito\footnote{richard.brito@tecnico.ulisboa.pt}$^{1}$, Vitor Cardoso\footnote{vitor.cardoso@tecnico.ulisboa.pt}$^{1,2}$, Paolo Pani\footnote{paolo.pani@tecnico.ulisboa.pt}$^{1,3}$}

\address{$^{1}$ CENTRA, Departamento de F\'{\i}sica, Instituto Superior T\'ecnico, Universidade de Lisboa,
Avenida Rovisco Pais 1, 1049 Lisboa, Portugal}
\address{$^{2}$ Perimeter Institute for Theoretical Physics Waterloo, Ontario N2J 2W9, Canada}
\address{$^{3}$ Dipartimento di Fisica, ``Sapienza'' Universit\`a di Roma, P.A. Moro 5, 00185, Roma, Italy}


\begin{abstract}
Superradiant instabilities of spinning black holes can be used to impose
strong constraints on ultralight bosons, thus turning black holes into effective particle detectors.
However, very little is known about the development of the instability and whether
its nonlinear time evolution accords to the linear intuition.
For the first time, we attack this problem by studying the impact of gravitational-wave emission and gas accretion on the evolution
of the instability. Our quasi-adiabatic, fully-relativistic analysis
shows that: (i) gravitational-wave emission does not have a significant effect on the evolution of the black hole, (ii) accretion plays an important role and (iii) although the mass of the scalar cloud developed through superradiance can be a sizeable fraction of the black-hole mass, its energy-density is very low and backreaction is negligible. Thus, massive black holes are well described by the Kerr geometry even if they develop bosonic clouds through superradiance.
Using Monte Carlo methods and very conservative assumptions, we provide strong support to the validity of the linearized analysis and to the bounds of previous studies. 
\end{abstract}




\maketitle


\section{Introduction}
One of the most solid predictions of Einstein's general relativity is that \emph{black holes (BHs) have no hair}~\cite{wheeler} and that all isolated, vacuum BHs in the Universe are described by the two-parameter Kerr family. Observing any deviation from this Kerr hypothesis --~a goal within the reach of upcoming gravitational-wave (GW)~\cite{Doeleman:2008qh,LIGO,VIRGO,KAGRA,ET,ELISA} and electromagnetic~\cite{Lu:2014zja,GRAVITY} facilities~-- would inevitably imply novel physics beyond general relativity.

It has been recently pointed out that stationary spinning BHs can develop ``hair'' in the presence of massive bosonic fields~\cite{Hod:2012px,Herdeiro:2014goa}. These new BH configurations exist at the threshold of the superradiant instability of the Kerr BH against massive scalar fields~\cite{Damour:1976kh,Detweiler:1980uk,Zouros:1979iw,Cardoso:2004nk,Cardoso:2013krh} (for a review on BH superradiance see~\cite{review}) and they can be interpreted as the nonlinear extension of linear bound states of frequency $\omega=m\Omega_H$, where $m$ is the azimuthal wave number and $\Omega_H$ is the angular velocity of the BH horizon. Such configurations require a complex scalar field, with time and azimuthal dependence $\sim e^{im(\varphi-\Omega_H t)}$ otherwise a net scalar flux at the horizon and GW flux at infinity would prevent the geometry from being stationary.
Formation scenarios for such configurations based on collapse or Jeans-like instability arguments are, notwithstanding, hard to devise.

\begin{figure}[ht]
\begin{center}
\begin{tabular}{c}
\epsfig{file=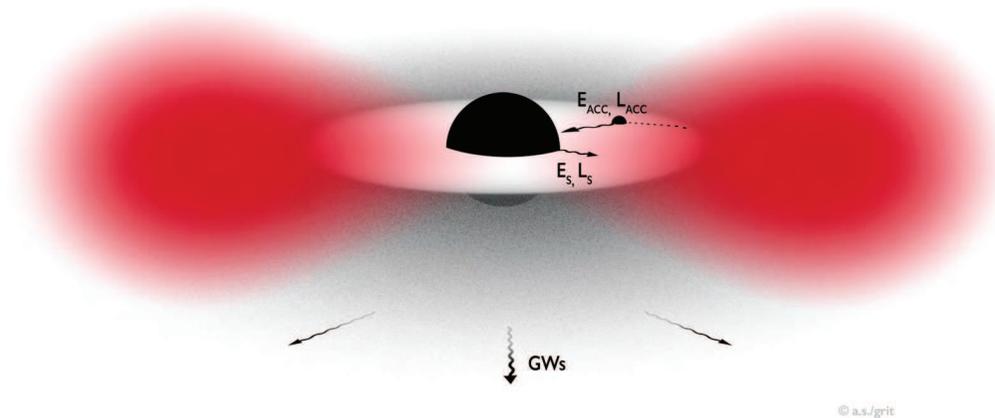,width=0.9\textwidth,angle=0,clip=true}
\end{tabular}
\end{center}
\caption{\label{fig:draw}
Pictorial description of a bosonic cloud around a spinning BH in a realistic astrophysical environment. The BH loses energy $E_S$ and angular momentum $L_S$ through superradiant extraction of scalar waves and emission of GWs, while accreting gas from the disk, which transports energy $E_{\rm ACC}$ and angular momentum $L_{\rm ACC}$. Notice that accreting material is basically in free fall 
after it reaches the innermost stable circular orbit. The cloud is localized at a distance $\sim 1/M\mu^2>2M$.}
\end{figure}
However, quantum or classical fluctuations of \emph{any} massive bosonic field trigger a superradiant instability of the Kerr metric, whose time scale $\tau$ can be extremely short. For a BH with mass $M$, the shortest instability time scale is $\tau\sim \left(\frac{M}{10^6 M_\odot}\right){\rm yr}$ for a ultralight scalar field~\cite{Cardoso:2005vk,Dolan:2007mj,Pani:2012vp,Witek:2012tr,Brito:2013wya}, and shorter for vector~\cite{Pani:2012vp,Pani:2012bp,Witek:2012tr} and tensor fields~\cite{Brito:2013wya} for which superradiance is more efficient.
Little is known about the nonlinear development of the instability but it is expected that, within such short time, a nonspherical bosonic cloud would grow near the BH extracting energy and angular momentum, until superradiance stops and the cloud is slowly re-absorbed by the BH and dissipated through GW emission~\cite{Arvanitaki:2010sy, Witek:2012tr,Okawa:2014nda,Cardoso:2013krh}. Although (at least for a real, stationary scalar field) the no-hair theorems~\cite{Hawking:1973uf,Heusler:1995qj,Sotiriou:2011dz,Graham:2014ina} guarantee that the final state of the instability has to be a Kerr BH with lower spin and no hair, it is important to understand the time scales involved in this process, because a scalar cloud surviving for cosmological times would be practically indistinguishable from a full-fledged BH hair and would have various important consequences.

A further motivation to explore realistic evolutions of the instability derives from the surprising connections between strong-field gravity and particle physics.
In recent years superradiant instabilities of astrophysical BHs have been used --~together with precision measurements of BH mass and spin (see e.g.~\cite{Brenneman:2011wz})~-- to constrain stringy axions and ultralight scalars~\cite{Arvanitaki:2010sy,Kodama:2011zc} (these constraints being complementary to those coming from cosmological observations~\cite{Bozek:2014uqa,Hlozek:2014lca}), to derive bounds on light vector fields~\cite{Pani:2012vp} and on the mass of the graviton~\cite{Brito:2013wya}, as well as to put intrinsic bounds on magnetic fields near BHs~\cite{Brito:2014nja} and on the fraction of primordial BHs in dark matter~\cite{Pani:2013hpa}. However, all these predictions are based on a linearized analysis, neglecting backreaction and other competitive effects --~such as GW emission and gas accretion~-- which can have an impact on the development of the process (see Fig.~\ref{fig:draw} for a pictorial view of the system under consideration). In this paper we take the first step to understand the evolution of the superradiant instability of a Kerr BH and to identify the relevant time scales for this problem.

Our main conclusion is that the linearized analysis used so far to impose constraints on particle masses is \emph{accurate} for both real and complex fields. In fact, a linearized analysis
of the instability might remain accurate in the entire regime of initial conditions, even when the mass of the scalar ``cloud'' that forms is of the order of the BH mass. The reason
is that the scalar field is typically distributed over a very large volume, implying very small densities and consequent small backreaction effects. For this reason, the BH geometry is very well described by the Kerr metric, even in the presence of massive bosonic clouds.

\section{Bosonic clouds around BHs: a quasi-adiabatic approximation \label{sec:formalism}}
For concreteness --~and also because it is where most of the work on BH superradiance is framed~-- we focus on the action for a minimally coupled massive scalar field, which can be either real or complex (we use Planck units):
\be
S=\int d^4x \sqrt{-g} \left( \frac{R}{16\pi}-\frac{1}{2}g^{\mu\nu}\Psi^{\ast}_{,\mu}\Psi^{}_{,\nu} - \frac{\mu^2}{2}\Psi^{\ast}\Psi\right)\,,\label{eq:MFaction}
\ee
although the qualitative aspects of our analysis are valid also for other massive bosonic fields\footnote{Here we neglect possible scalar self-interactions beyond the mass term. Nonlinearities can give rise to interesting effects, such as bosenova explosions in axion clouds~\cite{Yoshino:2012kn}, which occur when the energy of the axion field is comparable to the BH mass.}.
The resulting field equations are $\nabla_{\mu}\nabla^{\mu}\Psi =\mu^2\Psi$ and $G^{\mu\nu}=8\pi T^{\mu\nu}$ with
\begin{equation}
T^{\mu \nu}=\Psi^{\ast,(\mu}\Psi^{,\nu)}-\frac{1}{2}g^{\mu\nu}\left( \Psi^{\ast}_{,\alpha}\Psi^{,\alpha}+{{\mu}^2}\Psi^{\ast}\Psi\right)\,. \label{Tmunu}
\end{equation}
A full nonlinear evolution of this system in the case of a spinning BH was recently performed~\cite{Okawa:2014nda}; following the development of the instability is extremely challenging because of the time scales involved: 
$\tau_{\rm BH}\sim M$ is the light-crossing time, $\tau_S\sim 1/\mu$ is the typical oscillation period of the scalar cloud and $\tau \sim M/(M\mu)^9$ is the instability time scale in the small-$M\mu$ limit. In the most favorable case for the instability, $\tau\sim 10^6\tau_S$ is the minimum evolution time scale required for the superradiant effects to become noticeable\footnote{The minimum instability time scale corresponds to $M\mu\sim0.42$ (see e.g.~\cite{Dolan:2007mj}). Although this value is beyond the analytical, small-$M\mu$ approximation, the numerical result is in good agreement with an extrapolation of the analytical formula~\cite{Pani:2012bp}.}. Thus, current nonlinear evolutions (which typically last at most $\sim 10^3 \tau_S$~\cite{Okawa:2014nda}) have not yet probed
the development of the instability, nor the impact of GW emission.

However, in such configuration the system is suitable for a quasi-adiabatic approximation: over the dynamical time scale of the BH the scalar field can be considered almost stationary and its backreaction on the geometry can be neglected as long as the scalar energy is small compared to the BH mass.  
Therefore, we consider a perturbative expansion in powers of the scalar field and check consistency a posteriori.
\subsection{Linearized analysis}
At leading order, the geometry is described by the Kerr spacetime and the scalar evolves in this fixed background. In the Teukolsky formalism~\cite{Teukolsky:1972my,Teukolsky:1973ha}, the Klein-Gordon equation can be separated by use of spin-0 spheroidal wavefunctions~\cite{Berti:2005gp},
\be
\Psi=\int d\omega e^{-i\omega t+im\varphi}{_0}S_{lm}(\vartheta)\psi(r)\,, \nn
\ee
and is equivalent to the following differential equations,
\beq
{\cal D}_\vartheta[{_0}S]
&+&\left[a^2(\omega^2-\mu^2)\cos^2\vartheta
-\frac{m^2}{\sin^2\vartheta}+\lambda\right]{_0}S=0\,,\nn\\
{\cal D}_r[\psi]&+&\left[\omega^2(r^2+a^2)^2-4aMrm\omega+a^2m^2-\Delta(\mu^2r^2+a^2\omega^2+\lambda)\right]\psi=0\,.\nn
\eeq
where ${\cal D}_\vartheta=(\sin\vartheta)^{-1}\partial_\vartheta\left(\sin\vartheta\partial_\vartheta\right)$, ${\cal D}_r=\Delta\partial_r\left(\Delta\partial_r\right)$, $\Delta=(r-r_+)(r-r_-)$, $r_\pm=M\pm \sqrt{M^2-a^2}$ and $a$ is related to the BH angular momentum $J=a M$.
A numerical solution to the above coupled system is straightforward~\cite{Berti:2009kk,Cardoso:2005vk}. For small mass couplings $M\mu$, it can be shown that the corresponding eigenvalue problem admits a hydrogenic-like solution~\cite{Detweiler:1980uk,Pani:2012vp,Pani:2012bp} with $\lambda\sim l(l+1)$ and
\be\label{omega}
\omega\sim \mu-\frac{\mu}{2}\left(\frac{M\mu}{l+n+1}\right)^2+\frac{i}{\gamma_l}\left(\frac{am}{M}-2\mu r_+\right)(M\mu)^{4l+5}\,,
\ee
where $n=0,1,2...$ and $\gamma_1=48$ for the dominant unstable $l=1$ mode. Note that the eigenfrequencies are complex, $\omega=\omega_R+i\omega_I$, unless the superradiant condition is saturated, at 
\begin{equation}
 a=a_{\rm crit}\approx \frac{2\mu M r_+}{m}\,.\label{acrit}
\end{equation}

In the small-$\mu$ limit the eigenfunctions read~\cite{Detweiler:1980uk,Yoshino:2013ofa}
\be
\psi(\mu,a,M,r)=A_{ln} g(\tilde{r})\,,
\ee
where $g(\tilde{r})$ is an universal function of the dimensionless quantity $\tilde{r}=2r M {\mu}^2/(l+n+1)$ and can be written in terms of Laguerre polynomials
\begin{equation}
 g(\tilde r)=\tilde{r}^l e^{-\tilde{r}/2}L_{n}^{2l+1}(\tilde{r})\,.\label{univg}
\end{equation}
We have verified that this is a good description of the numerical eigenfunctions for moderately large $\mu M\lesssim0.2$ even at large BH spin. Notice that the eigenfunction peaks at $r_{\rm cloud}\sim \frac{(l+n+1)^2}{(M\mu)^2}M$~\cite{Arvanitaki:2010sy} (see also~\cite{Benone:2014ssa}) and thus extends way beyond the horizon, where rotation effects can be neglected. 
For definiteness, and because it is the single most unstable mode, we focus for now on $l=m=1$ and $n=0$. In this case $g(\tilde r)=\tilde{r} e^{-\tilde{r}/2}$.

As we noted, there are clearly two scales in the problem. One is dictated by the oscillation time $\tau_S=1/\omega_R\sim1/\mu$, the other by the instability growth time scale, $\tau=1/\omega_I\gg\tau_S$.
As such, we will consider these scales to be well separated, and will assume that the cloud is stationary and described by
\be
\Psi=A_0g(\tilde{r})\cos\left(\varphi-\omega_Rt\right)\sin\vartheta\,, \label{scalar}
\ee
where $A_0\equiv A_{10}$. In Eq.~\ref{scalar} we assumed a {\it real} scalar field, because this is the configuration that maximizes GW emission:
complex scalars in a nearly stationary regime will exhibit no time-dependent stress-energy tensor, and therefore do not emit GWs in this approximation (this case is briefly discussed in Sec.~\ref{sec:hairy} below).
As we will show, even real scalars give rise to very small GW energy fluxes.

For convenience, by using Eq.~\ref{Tmunu}, the amplitude $A_0$ can be expressed in terms of the mass $M_S$ of the scalar cloud,
\be
M_S=\int r^2\sin\vartheta\rho=\frac{2\pi A_0^2}{3} \left(2{\cal I}_0+\frac{2{\cal I}_2}{M^2{\mu}^2}+{\cal I}'_2\right)\,,
\ee
where we defined the dimensionless integrals 
\begin{eqnarray}
 {\cal I}_n&=& \int_0^\infty d\tilde{r} \tilde{r}^n g(\tilde{r})^2\,,\quad
 {\cal I}'_n= \int_0^\infty d\tilde{r} \tilde{r}^n g'(\tilde{r})^2\,,
\end{eqnarray}
and the energy density $\rho\equiv -T_0^0$ reads
\begin{eqnarray}
 \rho &=& \frac{A_0^2}{2r^2}\left\{{\mu}^4 M^2 r^2 \sin ^2(\vartheta ) g'(\tilde{r})^2 \cos ^2(\varphi -\omega_R t)\right.\nn\\
 &&\left.+g(\tilde{r})^2 \left[\left(\cos ^2(\vartheta )+{\mu}^2 r^2 \sin ^2(\vartheta )\right) \cos ^2(\varphi -\omega_R t)\right.\right.\nn\\
 &&\left.\left.+\left[1+r^2 \omega_R^2 \sin ^2(\vartheta )\right] \sin ^2(\varphi -\omega_R t)\right]\right\}\,, \label{rho}
\end{eqnarray}
where a prime denotes derivative with respect to the argument. 
In the small ${\mu}M$ limit one obtains
\begin{equation}\label{amplitude}
A_0^2=\frac{3}{4\pi {\cal I}_2}\left(\frac{M_S}{M}\right) ({\mu} M)^4\,. 
\end{equation}
In deriving the formulas above we have assumed that spacetime is flat. This approximation is accurate as long as the cloud is localized far away from the BH, i.e. when $\mu M\ll1$ (cf. Ref.~\cite{Yoshino:2013ofa} where a similar approximation is discussed). When $\mu M\ll 1$, the relation~\ref{amplitude} is valid also in the full Kerr case.
\subsection{GW emission}

A nonspherical monochromatic cloud as in Eq.~\ref{scalar} will emit GWs with frequency $ 2\pi/\lambda\sim 2\omega_R\sim 2\mu$, the wavelength $\lambda$ being in general \emph{smaller} than the size of the source, $r_{\rm cloud}$. Thus, even though the cloud is nonrelativistic, the quadrupole formula does not apply because the emission is incoherent~\cite{Arvanitaki:2010sy,Yoshino:2013ofa}. However, due to the separation of scales between the size of the cloud and the BH size for $\mu M\ll 1$, the GW emission can be analyzed taking the source to lie in a nonrotating (or even flat~\cite{Yoshino:2013ofa}) background.

In the fully relativistic regime, the gravitational radiation generated is best described by the Teukolsky formalism
for the gravitational perturbations. This is done in~\ref{sec:RWZ}. Our results, in agreement with a previous analysis by Yoshino and Kodama~\cite{Yoshino:2013ofa},
yield the following energy flux,
\be
\dot{E}_{\rm GW}=\frac{484+9 \pi ^2}{23040}\left(\frac{M_S^2}{M^2}\right)(M\mu)^{14}\,. \label{dEdtF}
\ee
The different prefactor relative to that derived in Ref.~\cite{Yoshino:2013ofa} is due to the fact that we considered a Schwarzschild background instead of a flat metric. Indeed, our result~\ref{dEdtF} is in better agreement with the numerical results.
Note that such flux is an \emph{upper bound} relative to the exact numerical results which are valid for any $\mu$ and any BH spin~\cite{Yoshino:2013ofa}. In the following we will use Eq.~\ref{dEdtF} as a very conservative assumption, since the GW flux is generically smaller.

A similar computation for the angular momentum dissipated in GWs gives
\be
\dot{J}_{\rm GW}=\frac{1}{\omega_R} \dot{E}_{\rm GW}\,, \label{dJdtF}
\ee
in agreement with the general result for a monochromatic wave of the form~\ref{scalar}.

\subsection{Accretion}
Astrophysical BHs are not in isolation but surrounded by matter fields in the form of gas and plasma. On the one hand, addition of mass and angular momentum to the BH via accretion competes with superradiant extraction. On the other hand, a slowly-rotating BH which does not satisfy the superradiance condition might be spun up by accretion and might become superradiantly unstable precisely \emph{because} of angular momentum accretion. Likewise, for a light BH whose coupling parameter $\mu M$ is small, superradiance might be initially negligible but it can become important as the mass of the BH grows through gas accretion. It is therefore crucial to include accretion in the treatment of BH superradiance, as we do here for the first time.

We make the most conservative assumption by using a model in which mass accretion occurs at a fraction of the Eddington rate (see e.g.~\cite{Barausse:2014tra}):
\begin{equation}
 \dot M_{\rm ACC} \equiv f_{\rm Edd} \dot M_{\rm Edd}\sim 0.02 f_{\rm Edd} \frac{M(t)}{10^6 M_\odot} M_\odot {\rm yr}^{-1}\,,\label{dotMaccr}
\end{equation}
where we have assumed an average value of the radiative efficiency $\eta\approx0.1$, as required by Soltan-type arguments, i.e. a comparison between the luminosity of active galactic nuclei and the mass function of BHs~\cite{LyndenBell:1969yx,Soltan:1982vf}. The Eddington ratio for mass accretion, $f_{\rm Edd}$, depends on the details of the accretion disk surrounding the BH and it is at most of the order unity for quasars and active galactic nuclei, whereas it is typically much smaller for quiescent galactic nuclei (e.g. $f_{\rm Edd}\sim 10^{-9}$ for SgrA$^{*}$). If we assume that mass growth occurs via accretion through Eq.~\ref{dotMaccr}, the
BH mass grows exponentially with $e$-folding time given by a fraction $1/f_{\rm Edd}$ of the Salpeter time scale, $\tau_{\rm Salpeter}=\frac{\sigma_T}{4\pi m_p}\sim4.5\times 10^7$~yr, where $\sigma_T$ is the Thompson cross section and $m_p$ is the proton mass. Therefore, the minimum time scale for the BH spin to grow via gas accretion is roughly $\tau_{\rm ACC}\sim \tau_{\rm Salpeter}/f_{\rm Edd}\gg \tau_{\rm BH}$ and also in this case the adiabatic approximation is well justified.

Regarding the evolution of the BH angular momentum through accretion, we make the conservative assumption that the disk lies on the equatorial plane and extends down to the innermost stable circular orbit (ISCO). 
If not, angular momentum increase via accretion is suppressed and superradiance becomes (even) more dominant.
Ignoring radiation effects, the evolution equation for the spin reads~\cite{Bardeen:1970zz}
\begin{equation}
\dot J_{\rm ACC} \equiv \frac{L(M,J)}{E(M,J)} \dot M_{\rm ACC}\,,\label{dotJaccr}
\end{equation}
where $L(M,J)=2M/(3\sqrt{3})\left(1+2 \sqrt{3 r_{\rm ISCO}/{M}-2}\right)$ and $E(M,J)=\sqrt{1-2M/3r_{\rm ISCO}}$ are the angular momentum and energy per unit mass, respectively, of the ISCO of the Kerr metric, located at $r_{\rm ISCO}=r_{\rm ISCO}(M,J)$ in Boyer-Lindquist coordinates. 

In the absence of superradiance the BH would reach extremality in finite time, whereas radiation effects set an upper bound of $a/M\sim 0.998$~\cite{Thorne:1974ve}. To mimic this upper bound in a simplistic way, we introduced a smooth cutoff in the accretion rate for the angular momentum. This cutoff merely prevents the BH to reach extremality and does not play any role in the evolution discussed in the next section.

\begin{figure*}[ht]
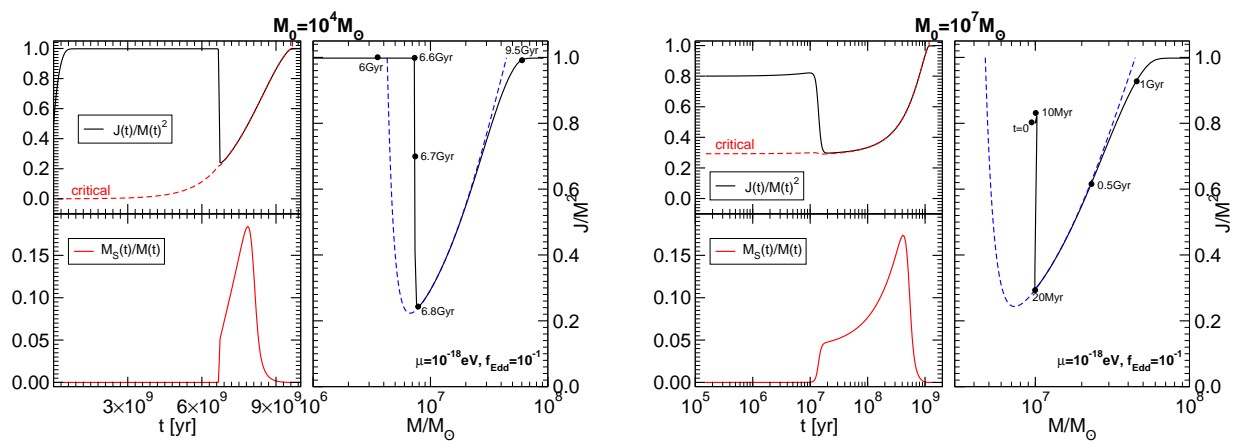

\begin{center}
\begin{tabular}{ccc}
\epsfig{file=evolution_M01e4_mu1em18_J05_fEdd_1em1.eps,width=0.49\textwidth,angle=0,clip=true}&\hspace{0.3cm}&
\epsfig{file=evolution_M01e7_mu1em18_J08_fEdd_1em1.eps,width=0.49\textwidth,angle=0,clip=true}
\end{tabular}
\end{center}
\caption{\label{fig:evolution}
Evolution of the BH mass and spin and of the scalar cloud due to superradiance, accretion of gas and emission of GWs. The two sets of plots show two different cases. In Case I (left set) the initial BH mass $M_0=10^4 M_\odot$ and the initial BH spin $J_0/M_0^2=0.5$. The BH enters the instability region at about $t\sim 6{\rm Gyr}$, when its mass $M\sim10^7 M_\odot$ and its spin is quasi-extremal. The set of plots on the right shows Case II, in which $M_0=10^7M_\odot$ and $J_0/M_0^2=0.8$, and the evolution starts already in the instability region for this scalar mass $\mu=10^{-18}{\rm eV}$. For both cases, the left top panels show the dimensionless angular momentum $J/M^2$ and the critical superradiant threshold $a_{\rm crit}/M$ (cf. Eq.~\ref{acrit}); the left bottom panels show the mass of the scalar cloud $M_S/M$ (note the logarithmic scale in the x-axis for Case II); and the right panels show the trajectory of the BH in the Regge plane~\cite{Arvanitaki:2010sy} during the evolution. The dashed blue line denotes the depleted region as estimated by the linearized analysis, i.e. it marks the threshold at which $\tau\sim\tau_{\rm ACC}$. 
}
\end{figure*}
%

\section{Evolution of the cloud}
We are now in a position to discuss the evolution of the scalar cloud within the quasiadiabatic approximation. 
The scalar energy flux that is extracted from the horizon through superradiance is
\be
\dot{E}_S=2M_S\omega_I\,,
\ee
where $M\omega_I=\frac{1}{48}({a/M-2{\mu} r_+})(M{\mu})^9$
for the $l=m=1$ fundamental mode and clearly $\dot{E}_S>0$ only in the superradiant regime.  

Two further contributions come from the emission of GWs through fluxes~\ref{dEdtF} and~\ref{dJdtF} and from gas accretion through the accretion rates~\ref{dotMaccr} and \ref{dotJaccr}. Energy and angular momentum conservation requires that
\begin{eqnarray}
 \dot{M}+\dot{M}_S&=&-\dot{E}_{\rm GW}+\dot M_{\rm ACC}\,,\\
 \dot{J}+\dot{J}_S&=&-\frac{1}{{\mu}}\dot{E}_{\rm GW}+\dot J_{\rm ACC}\,,
\end{eqnarray}
where we have used $\dot{J}_{\rm GW}=\dot{E}_{\rm GW}/\omega_R\sim\dot{E}_{\rm GW}/\mu$, we have neglected the subdominant contributions of the mass of the disk and of those GWs that are absorbed at the horizon, and we have approximated the local mass and angular momentum by their ADM counterparts. The latter approximation is valid as long as backreaction effects are small, as we discuss below.
The evolution of the system is governed by the two equations above supplemented by 
\beq
\dot{M}&=& -\dot{E}_S+\dot M_{\rm ACC} \,,\\
\dot{J}&=& -\frac{1}{{\mu}}\dot{E}_S +\dot J_{\rm ACC}\,,
\eeq
which describe the superradiant extraction of energy and angular momentum and the competitive effects of gas accretion at the BH horizon. These equations assume that the scalar cloud is not directly (or only very weakly)  coupled to the disk.

Representative results for the evolution of the system are presented in Fig.~\ref{fig:evolution} where we consider the scalar-field mass $\mu=10^{-18}{\rm eV}$ and mass accretion near the Eddington rate, $f_{\rm Edd}=0.1$. We consider two cases: (I) the left set of plots corresponds to a BH with initial mass $M_0=10^4 M_\odot$ and initial spin $J_0/M_0^2=0.5$, whereas (II) the right set of plots corresponds to $M_0=10^7 M_\odot$ and $J_0/M_0^2=0.8$.

In Case I, superradiance is initially negligible because $\mu M_0\sim 10^{-4}$ and superradiant extraction is suppressed. Thus, the system evolves mostly through gas accretion, reaching extremality ($J/M^2\sim0.998$) within the time scale $\tau_{\rm ACC}\sim 10\tau_{\rm Salpeter}$. At about $t\sim 6{\rm Gyr}$, the BH mass is sufficiently large that the superradiant coupling $\mu M$ becomes important. This corresponds to the BH entering the region delimited by a dashed blue curve in the Regge plane~\cite{Arvanitaki:2010sy} shown in Fig.~\ref{fig:evolution} for Case I. At this stage superradiance becomes effective very quickly: a scalar cloud grows exponentially near the BH (left bottom panel), while
mass and angular momentum are extracted from the BH (left top panel). This abrupt phase lasts until the BH spin reaches the critical value $a_{\rm crit}/M$ and superradiance halts. Because the initial growth is exponential, the evolution does not depend on the initial mass and initial spin of the scalar cloud as long as the latter are small enough, so that in principle also a quantum fluctuation would grow to a sizeable fraction of the BH mass in finite time. 

Before the formation of the scalar condensate, the evolution is the same regardless of GW emission and the only role of accretion is to bring the BH into the instability window.  After the scalar growth, the presence of GW dissipation and accretion produces two effects: (i) the scalar condensate loses energy through the emission of GWs, as shown in the left bottom panel of Fig.~\ref{fig:evolution}; (ii) gas accretion returns to increase the BH mass and spin.

However, because accretion restarts in a region in which the superradiance coupling $\mu M$ is nonnegligible, the ``Regge trajectory'' $J(t)/M(t)^2\sim a_{\rm crit}/M$ (cf. Eq.~\ref{acrit}) is an attractor for the evolution and the BH ``stays on track'' as its mass and angular momentum grow. For Case I, this happens between $t\sim 6.8{\rm Gyr}$ and $t\sim9.5{\rm Gyr}$, i.e. until the spin reaches the critical value $J/M^2\sim 0.998$ and angular momentum accretion saturates.

A similar discussion holds true also for Case II, presented in the right set of plots in Fig.~\ref{fig:evolution}. In this case, the BH starts already in the instability regime, its spin grows only very little before superradiance becomes dominant, and the BH angular momentum is extracted in about $10{\rm Myr}$. After superradiant extraction, the BH evolution tracks the critical value $a_{\rm crit}/M$ while the BH accretes over a time scale of $1{\rm Gyr}$.

\subsection{The role of accretion}

While GW emission is always too weak to affect the evolution of the BH mass and spin (nonetheless being responsible for the decay of the scalar condensate as shown in Fig.~\ref{fig:evolution}), accretion 
plays a more important role. From Fig.~\ref{fig:evolution}, it is clear that accretion produces two effects. First, for BHs which initially are not massive enough to be in the superradiant instability region, accretion brings them to the instability window by feeding them mass as in Case I. Furthermore, when $J/M^2\to a_{\rm crit}/M$ the superradiant instability is exhausted, so that accretion is the only relevant process and the BH inevitably spins up again. This accretion phase occurs in a very peculiar way, with the dimensionless angular momentum following the trajectory $J/M^2\sim a_{\rm crit}/M$ over very long time scales.

Therefore, a very solid prediction of BH superradiance is that supermassive BHs would move on the Regge plane following the bottom-right part of the superradiance threshold curve. The details of this process depend on the initial BH mass and spin, on the scalar mass $\mu$ and on the accretion rate.

Thus, in order to verify the theoretical bounds on the existence of light bosons~\cite{Arvanitaki:2010sy,Kodama:2011zc,Pani:2012vp,Brito:2013wya}, a relevant problem concerns the \emph{final} BH state at the time of observation. In other words, given the observation of an old BH and the measurement of its mass and spin, would these measurements be compatible with the evolution depicted in Fig.~\ref{fig:evolution}? 

\begin{figure}[ht]
\begin{center}
\begin{tabular}{c}
\epsfig{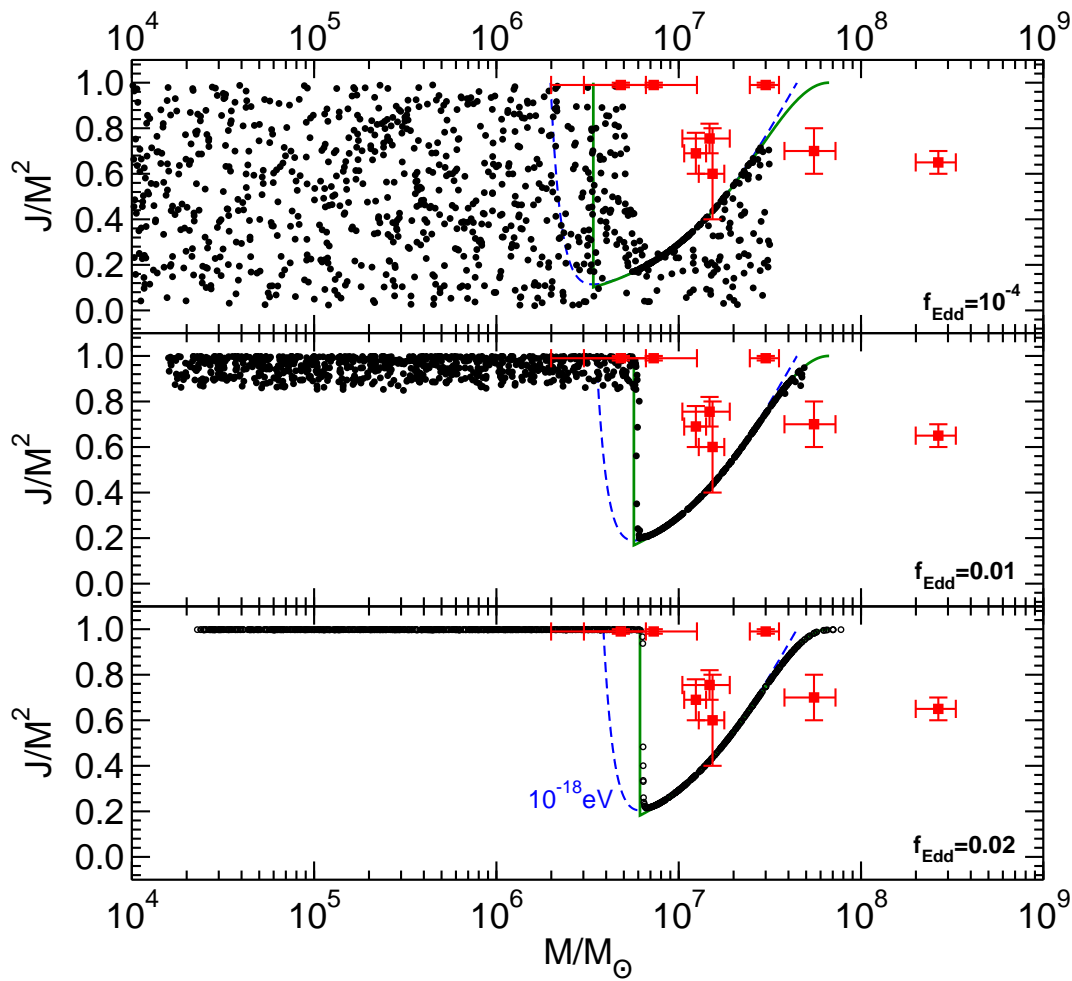}
\end{tabular}
\end{center}
\caption{\label{fig:ReggeMC}
The final BH mass and spin in the Regge plane for initial data consisting of $N=10^3$ BHs with initial mass and spin randomly distributed between $\log_{10}M_0\in[4,7.5]$ and $J_0/M_0^2\in[0.001,0.99]$. The BH parameters are then extracted at $t=t_F$, where $t_F$ is distributed on a Gaussian centered at $\bar t_{F}\sim 2\times 10^9{\rm yr}$ with width $\sigma=0.1\bar t_{F}$. We considered $\mu=10^{-18}{\rm eV}$. The dashed blue line is the prediction of the linearized analysis obtained by comparing the superradiant instability time scale with the accretion time scale, $\tau\approx\tau_{\rm Salpeter}/f_{\rm Edd}$, whereas the solid green line denotes the region defined through Eq.~\ref{region}. Old BHs do not populate the region above the green threshold curve. The experimental points with error bars refer to the supermassive BHs listed in~\cite{Brenneman:2011wz}.
}
\end{figure}

To assess this question, we have used Monte Carlo methods.
In Fig.~\ref{fig:ReggeMC} we show the final BH mass and spin in the Regge plane for $N=10^3$ evolutions for a scalar field mass $\mu=10^{-18}{\rm eV}$. These were obtained with random distributions of the initial BH mass between $\log_{10}M_0\in[4,7.5]$ and $J_0/M_0^2\in[0.001,0.99]$ extracted at $t=t_F$, where $t_F$ is distributed on a Gaussian centered at $\bar t_{F}\sim 2\times 10^9{\rm yr}$ with width $\sigma=0.1\bar t_{F}$. We consider three different accretion rates and, in each panel, we superimpose the bounds derived from the linearized analysis, i.e. the threshold line when the instability time scale equals the accretion time scale, $\tau\sim \tau_{\rm ACC}$. As a comparison, in the same plot we include the experimental points for the measured mass and spin of some supermassive BHs listed in Ref.~\cite{Brenneman:2011wz}.

Various comments are in order. First, it is clear that the higher the accretion rate the better the agreement with the linearized analysis. This seemingly counter-intuitive result can be understood by the fact that higher rates of accretion make it more likely to find BHs that have undergone a superradiant instability phase over our observational time scales. In fact, for high accretion rates it is very likely to find supermassive BHs precisely on the ``Regge trajectory''~\cite{Arvanitaki:2010sy} given by $J/M^2\sim a_{\rm crit}/M$ (cf. Eq.~\ref{acrit}).

Furthermore, for any value of the accretion rate, we always observe a depleted region (a ``hole'') in the Regge plane~\cite{Arvanitaki:2010sy}, which is not populated by old BHs. While the details of the simulations might depend on the distribution of initial mass and spin, the qualitative result is very solid and is a generic feature of the evolution. For the representative value $\mu=10^{-18}{\rm eV}$ adopted here, the depleted region is incompatible with observations~\cite{Brenneman:2011wz}. Similar results would apply for different values\footnote{Note that, through Eq.~\ref{dotMaccr}, the mass accretion rate only depends on the combination $f_{\rm Edd} M$, so that a BH with mass $M=10^6 M_\odot$ and $f_{\rm Edd}\sim10^{-3}$ would have the same accretion rate of a smaller BH with $M=10^{4} M_\odot$ accreting at rate $f_{\rm Edd}\sim10^{-1}$. Because this is the only relevant scale for a fixed value of $\mu M$, in our model the evolution of a BH with different mass can be obtained from Fig.~\ref{fig:evolution} by rescaling $f_{\rm Edd}$ and $\mu$.} of $\mu$ in a BH mass range such that $\mu M\lesssim1$. Therefore, as discussed in Refs.~\cite{Arvanitaki:2010sy,Pani:2012vp,Brito:2013wya}, observations of massive BHs with various masses can be used to rule out various ranges of the boson-field mass $\mu$.

Finally, Fig.~\ref{fig:ReggeMC} suggests that when accretion and GW emission are properly taken into account, the holes in the Regge plane are smaller than what naively predicted by the relation $\tau\approx\tau_{\rm ACC}$, i.e. by the dashed blue curve in Fig.~\ref{fig:ReggeMC}. Indeed, we find that a better approximation for the depleted region is 
\begin{equation}
 \frac{J}{M^2}\gtrsim \frac{a_{\rm crit}}{M}\sim 4\mu M \quad  \cup \quad M \gtrsim \left({\frac{96}{\mu ^{10} \tau_{\rm ACC}}}\right)^{1/9} \,,\label{region}
\end{equation}
whose boundaries are shown in Fig.~\ref{fig:ReggeMC} by a solid green line. These boundaries correspond to the threshold value $a_{\rm crit}$ (cf. Eq.~\ref{acrit}) for superradiance and to a BH mass which minimizes the spin for which $\tau\approx \tau_{\rm ACC}$, for a given $\mu$~\cite{Pani:2012bp}. As shown in Fig.~\ref{fig:ReggeMC}, the probability that a BH populates this region is strongly suppressed as the accretion rate increases.

\subsection{Estimating backreaction effects}
Our analysis neglects the gravitational effects of the scalar cloud and of a putative accretion disk on the BH geometry. The latter assumption is well justified because the disk density profile is roughly (see e.g.~\cite{Barausse:2014tra}):
\begin{equation}
 \frac{\rho_{\rm disk}}{\mbox{ kg}/\mbox{m}^3}\approx \left\{\begin{array}{l}
                           3.4\times 10^{-6} \left(\frac{10^6 M_\odot}{M}\right) \frac{f_{\rm Edd}}{\tilde{r}^{3/2}} \\
169 \frac{f_{\rm Edd}^{11/20}}{\tilde{r}^{15/8}} \left(1-\sqrt{\frac{\tilde{r}_{\rm in}}{\tilde{r}}}\right)^{11/20} \left(\frac{10^6 M_\odot}{M}\right)^{7/10} 
                          \end{array}\right.\nn \,,
\end{equation}
for geometrically-thick disks and for thin disks, respectively, where $\tilde{r}=r/M$ and $\tilde{r}_{\rm in}\sim 6$ is the radius of the inner edge of the disk in gravitational radii. These densities are negligible relative to the typical energy-density of the BH, $1/M^2\sim 10^8  (10^6 M_\odot/M)^2 {\rm kg/m}^3$, so that the deformation of the geometry due to the presence of the disk is unimportant.

\begin{figure}[ht]
\begin{center}
\begin{tabular}{c}
\epsfig{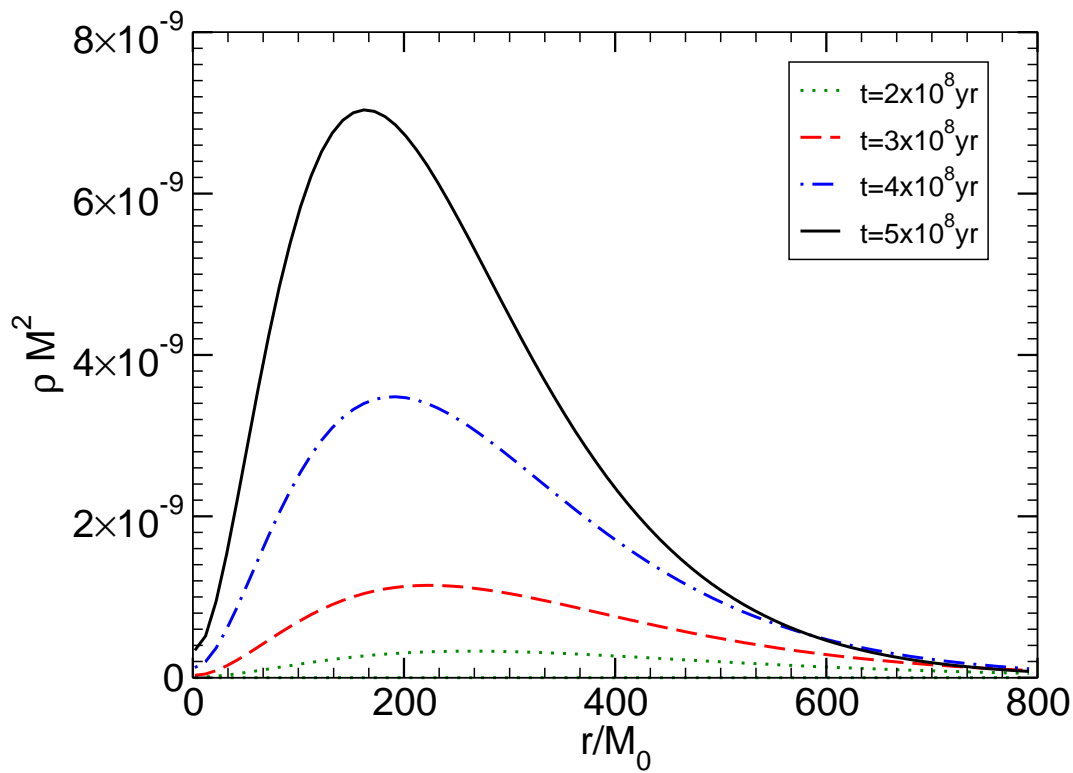}
\end{tabular}
\end{center}
\caption{\label{fig:density}
The energy-density profile of the scalar cloud on the equatorial plane and at azimuthal angle $\varphi=0$ in units of the BH density, $1/M^2\sim 10^6 {\rm kg/m}^3$, at different time snapshots for the evolution shown in the right panel of Fig.~\ref{fig:evolution}.
}
\end{figure}

On the other hand, from the evolution of Fig.~\ref{fig:evolution} it is clear that the scalar cloud attains a sizeable fraction of the total BH mass, so that backreaction effects might be relevant in this case. However, the scalar energy $M_S$ is spread over a large volume because the cloud peaks at $r_{\rm cloud}\sim \frac{1}{(M\mu)^2}M$. Thus, the scalar density --~which is the quantity directly coupled to the geometry through Einstein's equations~-- is always negligible. Figure~\ref{fig:density} shows the energy-density profile of the scalar cloud during the evolution corresponding to the right panel of Fig.~\ref{fig:evolution}. As in the case of the disk, also the density of the scalar cloud is orders of magnitude smaller than the energy-density associated to the BH horizon, $\sim 1/M^2$, so that the gravitational pull of the cloud produces a negligible effect on the background geometry. Furthermore, the corrections vanish near the BH horizon, so that also the superradiant energy extraction is unaffected\footnote{In principle, superradiant extraction from the BH horizon could be also affected by external perturbers, e.g. other compact objects in the vicinity of the BH. While our analysis already indicates that this correction is negligible, the fact that the near-horizon geometry of a BH cannot be easily deformed by, e.g., tidal forces~\cite{Binnington:2009bb} gives further support that considering isolated BHs in the context of superradiant instabilities is a reliable approximation (see also Ref.~\cite{Barausse:2014tra} for an analysis of environmental effects in GW physics).}.

This discussion is in agreement with the results obtained for the superradiant instability in modified Kerr geometries, for example the Kerr-de Sitter metric studied in Ref.~\cite{Zhang:2014kna}. In such case, a value of the cosmological constant comparable to that of the cloud density, $\Lambda\sim 10^{-8}/M^2$, has no impact on the instability. 
\subsection{Hairy BHs: do they ever form?}\label{sec:hairy}
The most plausible formation scenarios for BHs involve gravitational collapse of matter, and are likely to form -- on free-fall time scales -- a geometry which is well-described
by the Kerr metric. Our results then have two important consequences: the BH evolves through accretion, GW emission and superradiance,
but at late times it will not populate the region defined by Eq.~\ref{region}. In addition, the backreaction of the scalar condensate on the geometry is always small, i.e.,
even in the presence of a scalar cloud the geometry is that of a Kerr BH to very good approximation. This is relevant for electromagnetic tests of the Kerr hypothesis, which are ultimately based on geodesic motion and would likely not be able to detect the effects of the cloud directly. On the other hand, during the evolution the system emits a nearly monochromatic GW signal, which is an interesting source for next-generation GW detectors~\cite{Arvanitaki:2010sy,Yoshino:2012kn,Yoshino:2013ofa}.

Our results apply equally to real and complex scalars, and despite a recent work finding hairy BH solutions which depart significantly from the Kerr metric~\cite{Herdeiro:2014goa}.
The reason is that our formation scenario starts from a Kerr BH. Thus superradiance can only extract a finite amount of mass from the BH (in fact, at most 29\% of the initial BH mass, cf. e.g.~\cite{Begelman:2014aea}),
and therefore can only grow to a limited value. We find that this value is never sufficient to impart significant changes to the geometry.
By contrast, Refs.~\cite{Herdeiro:2014goa,Herdeiro:2014jaa} find that generically, stationary hairy BHs are smoothly connected to boson stars, and that therefore arbitrarily small BHs (or arbitrarily large ratios $M_S/M$)
are possible. What our results show is that these configurations do not arise from the evolution of initially isolated Kerr BHs; however, we have not ruled out that such solutions -- representing observationally large deviations from the Kerr geometry -- may arise as the end-state of some other initial conditions, most likely involving a large scalar field environment.

\begin{figure}[ht]
\begin{center}
\begin{tabular}{c}
\epsfig{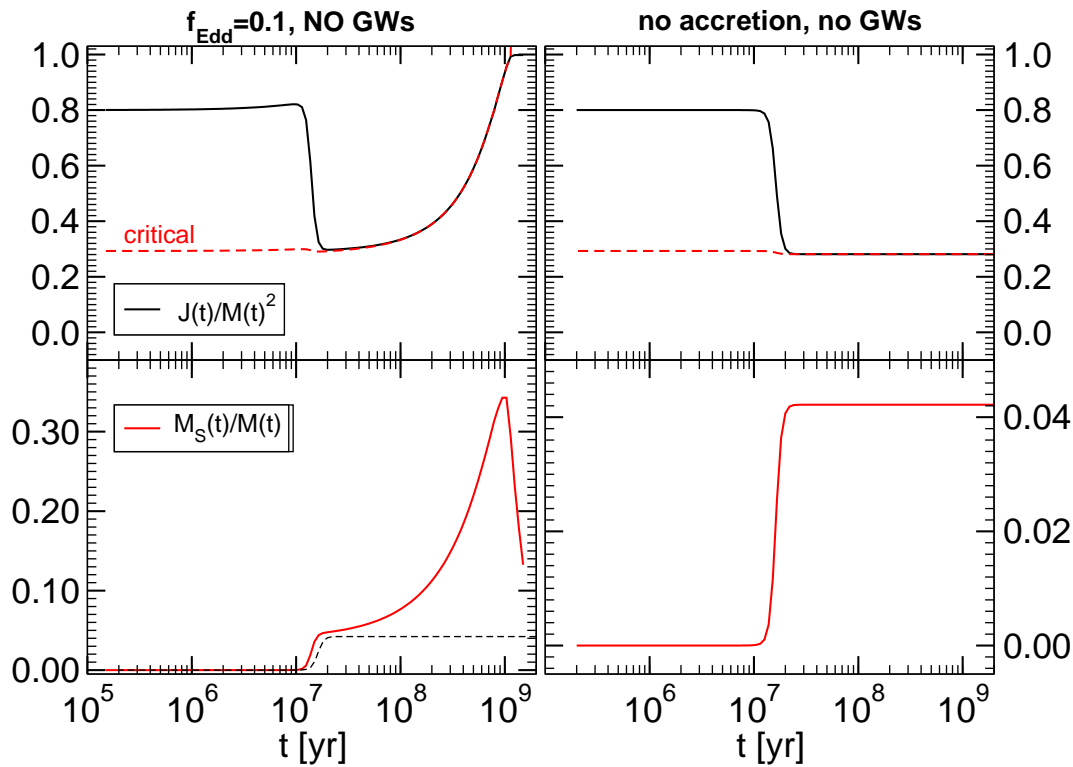}
\end{tabular}
\end{center}
\caption{\label{fig:NOGWs}
Evolution with the same initial conditions as in the right set of Fig.~\ref{fig:evolution} but turning off the emission of GWs as in the case of a complex scalar field. The left panels show the case of mass accretion at the rate $f_{\rm Edd}=0.1$, whereas the right panels show the case in which also accretion is turned off. For comparison, the scalar mass in the right bottom panel is also shown in the left bottom panel by a dashed black curve. When accretion is effective, the scalar cloud can become heavier.
}
\end{figure}

For completeness, in Fig.~\ref{fig:NOGWs} we show an evolution starting with the same initial conditions as in the right panel of Fig.~\ref{fig:evolution} but turning off GW emission, which corresponds to taking a stationary, complex-scalar cloud in place of Eq.~\ref{scalar}. In the left panels we consider the case of mass accretion at the rate $f_{\rm Edd}=0.1$, whereas in the right panels also accretion has been turned off. In the latter case, the scalar mass and the BH angular momentum saturate after the superradiant extraction and the system would never leave the plateau configuration with a scalar mass $M_S\sim 0.04 M$ and a reduced spin $J/M^2\sim0.3$. However, when accretion is turned on (left panels), the scalar mass can attain more than $30\%$ of the BH mass during the evolution. This is due to the fact that the BH mass and angular momentum grow through accretion when superradiance is still effective and can therefore continue feeding the scalar cloud. This process lasts until angular-momentum accretion becomes inefficient at $J/M^2\sim 0.998$. Nonetheless, even in this most favorable case for the growth of the scalar cloud, the energy-density of the scalar field is negligible and the geometry is very well described by the Kerr metric.

%

\subsection{Higher-$l$ modes}

So far we have neglected the superradiant growth of higher multipoles with $l>1$. This is justified by the fact that the instability time scale increases with $l$ (cf. Eq.~\ref{omega}) and the emission of GWs for increasing $l$ becomes even more negligible~\cite{Yoshino:2013ofa}. The spin-down process is always dominated by the lowest possible superradiant mode $l=m=1$. For example, for the  evolution shown in Fig.~\ref{fig:evolution} the spin down due to the growth of the multipole $l=2$ would only affect the evolution of the BH after time scales of the order $t\sim 10^{13}{\rm yr}$.

However, because the superradiance condition depends on the azimuthal number $m$, for certain parameters it might occur that the modes with $l=m=1$ are stable, whereas the modes with $l=m=2$ are unstable, possibly with a superradiant extraction stronger than accretion. When this is the case, our previous analysis confirms that the depleted region of the Regge plane is the union of various holes, as predicted in Ref.~\cite{Arvanitaki:2010sy} by using a linearized analysis.

In the case where axion nonlinearities are taken into account further spin-down due to higher multipoles is expected to be damped due to the axion self-interactions either through the mixing of superradiant with nonsuperradiant levels or through the occurrence of explosive nonlinear effects, such as the bosenova collapse of the axion cloud~\cite{Arvanitaki:2010sy,Yoshino:2012kn}.

\section{Discussion}
If ultralight bosonic degrees of freedom exist in nature, massive BHs should have a maximum spin lower than the Kerr bound and should be endowed with large dipolar bosonic clouds. Thus, observations of highly-spinning BHs can be used to constrain such fields, for example to put bounds on axions or on massive gravitons. Such predictions are based on a linearized analysis which neglects the effects of backreaction and other competitive effects such as accretion. In this paper, we have extended such analysis by including the emission of GWs from the cloud and the most conservative case of gas accretion. By adopting an adiabatic approximation, we have simulated the evolution of the scalar condensate around a spinning BH. 

Our results show that the effects of GW emission are always too small to affect the evolution of the BH mass and spin, but they contribute to dissipate the scalar condensate. Indeed, we have shown that the scalar condensates are eventually re-absorbed by the BH and dissipated through quadrupolar GWs, in accord to the BH no-hair theorems~\cite{Hawking:1973uf,Heusler:1995qj,Sotiriou:2011dz,Graham:2014ina}. Nonetheless, the mass of the cloud remains a sizeable fraction of the BH total mass over cosmological times, so that such systems can be considered as (quasi)-stationary hairy BHs for any astrophysical purpose. 
The energy-density in the scalar field is negligible and the geometry is very well described by the Kerr metric during the entire evolution. Thus, the prospects of imagining deviations from Kerr due to superradiantly-produced bosonic clouds in the electromagnetic band~\cite{Lu:2014zja,GRAVITY} are low, but such systems are a primary source for observations aiming at testing the Kerr hypothesis through GW detection~\cite{LIGO,VIRGO,KAGRA,ET,ELISA}.

The role of gas accretion is twofold. On the one hand, accretion competes against superradiant extraction of mass and angular momentum. On the other hand accretion might produce the optimal conditions for superradiance, for example by increasing the BH spin before the instability becomes effective or by ``pushing'' the BH into the instability region in the Regge plane.

Our Monte Carlo simulations confirm that a very generic prediction of BH superradiant instabilities is the existence of holes in the Regge plane. For mass accretion near the Eddington rate, such depleted regions are very well described by Eq.~\ref{region}, which corrects the estimate obtained just by comparing the instability time scale against a typical accretion time scale. A more sophisticated analysis --~including radiative effects and the geometry of the disk~-- would be important to refine the bounds previously derived~\cite{Arvanitaki:2010sy,Kodama:2011zc,Pani:2012vp,Brito:2013wya}.

The main limitation of our analysis is the assumption of adiabatic evolution, which is nonetheless well motivated given the large difference in the time scales of the problem. For the same reason, exact numerical simulations --~although important to test our results~-- would be extremely difficult to perform.  Some of our formulas were derived in the small $\mu M$ limit. Although they provide reliable results also when $\mu M\sim {\cal O}(1)$, our analysis could be extended using the exact numerical results derived in Ref.~\cite{Yoshino:2013ofa}. 
It would also be interesting to include scalar self-interactions, which are relevant for axions~\cite{Yoshino:2012kn}. Assuming the axion sine-Gordon potential would not change the evolution shown in Fig.~\ref{fig:evolution} considerably, our results suggest that the scalar cloud might reach the threshold ($M_S/M\gtrsim 0.16$) for the bosenova condensation~\cite{Yoshino:2012kn}.

Finally, we have focused on the scalar case but it is likely that similar results can be derived also for massive vector and tensor fields because in such cases the superradiant instability is stronger.

\vspace{0.1cm}
\noindent
{\bf Acknowledgments.}
We are indebted to Roberto Emparan, Carlos Herdeiro and Hirotaka Yoshino for useful comments and discussions.
We thank Ana Sousa for kindly preparing Figure~1 for us.
R.B. acknowledges financial support from the FCT-IDPASC program through the Grant No. SFRH/BD/52047/2012, and from the Funda\c c\~ao Calouste Gulbenkian through
the Programa Gulbenkian de Est\' imulo \`a Investiga\c c\~ao Cient\'ifica.
V.C. acknowledges financial support provided under the European Union's FP7 ERC Starting
Grant ``The dynamics of black holes: testing the limits of Einstein's theory''
grant agreement no. DyBHo--256667. 
This research was supported in part by the Perimeter Institute for Theoretical Physics. 
Research at Perimeter Institute is supported by the Government of Canada through 
Industry Canada and by the Province of Ontario through the Ministry of Economic Development 
$\&$ Innovation.
P.P. was supported by the European Community through
the Intra-European Marie Curie Contract No.~AstroGRAphy-2013-623439 and by FCT-Portugal through the projects IF/00293/2013 and CERN/FP/123593/2011.
This work was supported by the NRHEP 295189 FP7-PEOPLE-2011-IRSES Grant.
%

\appendix

\section{Gravitational wave emission\label{sec:RWZ}}

In the Teukolsky formalism the perturbation equations can be reduced to a second-order differential
equation for the Newman-Penrose scalar $\psi_4$. We can decompose $\psi_4$ as
\be
\psi_4(t,r,\Omega)=\sum_{lm} r^{-4}\int^\infty_{-\infty}d\omega\sum_{lm}R_{lm}(r)~_{-2}Y_{lm}(\Omega)e^{-i\omega t}\,, \label{psi4expansion}
\ee
where ${_s}Y_{lm}(\vartheta,\varphi)$ are the spin-$s$ weighted spherical harmonics~\cite{Berti:2005gp}. The radial function $R(r)$ satisfies the inhomogeneous equation
\beq\label{teu_eq}
&&r^2f R''-2(r-M)R'+\left[f^{-1}\left(\omega^2r^2-4i\omega (r-3M)\right)\right.\nn\\
&&\left.-(l-1)(l+2)\right]R=-T_{lm\omega}\,,
\eeq
where $f=1-2M/r$. The source term $T_{lm\omega}$ is related to the scalar field stress-energy tensor $T_{\mu\nu}$ through the tetrad projections, $T_{\mu\nu}n^{\mu}n^{\nu}\equiv T_{nn}$, $T_{\mu\nu}n^{\mu}\bar{m}^{\nu}\equiv T_{n\bar{m}}$ and $T_{\mu\nu}\bar{m}^{\mu}\bar{m}^{\nu}\equiv T_{\bar{m}\bar{m}}$, where
\beq
n^{\mu}&=&\frac{1}{2}\left (1,-f,0,0 \right )\,,\\
\bar{m}^{\mu}&=&\frac{1}{\sqrt{2}\,r}\left (0,0,1,-\frac{i}{\sin\vartheta}
\right )\,. \eeq 
We define
\beq 
_{S}T&\equiv& \frac{1}{2\pi} \int\, d\Omega\, dt \, {\cal T}_{S}~_{S}\bar{Y}_{lm}e^{i\omega t}\,,
\eeq
where ${\cal T}_S=T_{nn}$, $T_{n\bar{m}}$ and $T_{\bar{m}\bar{m}}$ for $S=0,-1,-2$, respectively. The source is then given by~\cite{Poisson:1993vp}
\begin{eqnarray}
\frac{T_{lm\omega}}{2\pi}&=&2\left[(l-1)l(l+1)(l+2)\right]^{1/2}r^4~_{0}T\nn\\
&+&2\left[2(l-1)(l+2)\right]^{1/2}r^2 f \mathcal{L}\left(r^3 f^{-1}~_{-1}T\right)\nn\\
&+&r f\mathcal{L}\left[r^4 f^{-1}\mathcal{L}\left(r~_{-2}T\right)\right]\,,
\end{eqnarray}
where $\mathcal{L}\equiv f \partial_r+i\omega$. Using Eqs.~\ref{Tmunu} and~\ref{scalar} we find that,
as expected, for the scalar configuration~\ref{scalar} the only modes that contribute are $l=|m|=2$ with frequencies $\omega=\pm 2\omega_R$.

Once the source term is known, $\psi_4$ can be computed using a Green's function approach. To construct the Green function we need two linearly independent solutions of the homogeneous equation. A physically motivated choice is to consider the solution $R^{\infty}$ which describes outgoing waves at infinity and $R^H$ which describes ingoing waves at the event horizon. By making use of the fact that the Wronskian $W=\frac{\partial_r R^{\infty} R^{H}-\partial_r R^H R^{\infty} }{r^2 f}=2i \omega B_{\rm in}$ is constant by virtue of the field equations, the correct solution of the inhomogeneous problem at infinity reads
\begin{equation}
R(r\to \infty)\sim \frac{R^{\infty}}{2i\omega B_{\rm in}}\int_{2M}^{\infty} dr \frac{R^H T_{lm\omega}}{r^4 f^2}\,, \label{sol}
\end{equation}
where $R^{\infty}(r\to \infty)\sim r^3 e^{i\omega r}$ and $R^H(r\to \infty)\sim B_{\rm out} r^3 e^{i\omega r}+B_{\rm in} e^{-i\omega r}/r$. From the asymptotic solution of Eq.~\ref{teu_eq}, we find
\be
B_{\rm in}=-\frac{C_1}{8\omega^2}(l-1)l(l+1)(l+2)e^{i(l+1)\pi/2}\,,
\ee
where $C_1$ is a constant. The solution $R^H$ can be found more easily by solving the Regge-Wheeler equation (see e.g.~\cite{Poisson:1993vp}) for small frequencies and using the fact that $R^H=r^2f\mathcal{L}\left(f^{-1}\mathcal{L}r \psi^H\right)$, where $\psi^H$ is the Regge-Wheeler function that, at small frequencies, reads~\cite{Poisson:1993vp}
\be 
\psi^H\sim C_1\omega r j_l(\omega r)\,,
\ee
where $j_l$ denote the spherical Bessel functions of the first kind.
Finally, the luminosity can be computed from 
\be
\dot{E}_{\rm GW}=\int d\Omega d\omega\frac{r^2}{4\pi\omega^2}|\psi_4|^2=\int d\omega\frac{|Z|^2}{2\pi\omega^2|W|^2}\,,
\ee
where $Z\equiv \int dr \frac{R^H T_{22\omega}}{r^4 f^2}$ and we used the fact that the modes with $m=\pm 2$ give the same contribution to the luminosity. 
The final result reads
\be
\dot{E}_{\rm GW}=\frac{2}{45} \pi ^2 \left(484+9 \pi ^2\right) M^6 A_0^4 \mu^6\,, \label{dEdtF2}
\ee
which, by using Eq.~\ref{amplitude} and $\mathcal{I}_2\sim 24$, reduces to the one quoted in the main text in the small-$\mu$ limit.

\vspace{-0.2cm}
\section*{References}
\bibliography{cloud}
\bibliographystyle{iopart-num}
\end{document}